\documentclass{pasj00}
\begin{document}
\SetRunningHead{R. Kitai et al.}{Umbral Fine Structures in Sunspots}
\Received{2000/12/31}%{yyyy/mm/dd}
\Accepted{2001/01/01}%{yyyy/mm/dd}

\title{Umbral Fine Structures in Sunspots Observed with Hinode Solar Optical Telescope}

%%% begin:list of authors
% Do NOT capitalize all letters in "textsc".
\author{
Reizaburo \textsc{Kitai}\altaffilmark{1},
Hiroko \textsc{Watanabe}\altaffilmark{1},
Tahei \textsc{Nakamura}\altaffilmark{1},
Ken-ichi \textsc{Otsuji}\altaffilmark{1},
Takuma \textsc{Matsumoto}\altaffilmark{1},
Satoru \textsc{UeNo}\altaffilmark{1},
Shin'ichi \textsc{Nagata}\altaffilmark{1},
Kazunari \textsc{Shibata}\altaffilmark{1},
Richard \textsc{Muller}\altaffilmark{2},
Kiyoshi \textsc{Ichimoto}\altaffilmark{3},
Saku \textsc{Tsuneta }\altaffilmark{3},
Yoshinori \textsc{Suematsu}\altaffilmark{3},
Yukio \textsc{Katsukawa}\altaffilmark{3},
Toshifumi \textsc{Shimizu}\altaffilmark{4},
Theodore D. \textsc{Tarbell}\altaffilmark{5},
Richard A. \textsc{Shine}\altaffilmark{5},
Alan M. \textsc{Title}\altaffilmark{5},
Bruce W. \textsc{Lites}\altaffilmark{6}
 } %
\altaffiltext{1}{ Kwasan and Hida Observatory, Kyoto University, Kamitakara, Gifu 506-1314, JAPAN }
\altaffiltext{2}{ Midi-Pyr\`en\`ees Observatory, France }
\altaffiltext{3}{ National Astronomical Observatory of Japan, Mitaka, Japan }
\altaffiltext{4}{ Institute of Space and Astronautical Science, JAXA, Sagamihara, Japan }
\altaffiltext{5}{ Lockheed Martin Solar and Astrophysics Laboratory, USA}
\altaffiltext{6}{ High Altitude Observatory, National Center for Atmospheric Research, USA}

\email{kitai@kwasan.kyoto-u.ac.jp}

%% `\KeyWords{}' always has to be placed before `\maketitle'.
\KeyWords{Sun:sunspot --- Sun:umbral dots --- Sun:magnetoconvection} %Do NOT move this preamble from here!

\maketitle

\begin{abstract}
High resolution imaging observation of a sunspot umbra was done with Hinode Solar Optical Telescope (SOT). Filtergrams in wavelengths of blue and green continuum were taken during three consecutive days. The umbra consisted of a dark core region, several diffuse components and numerous umbral dots. We derived basic properties of umbral dots (UDs), especially their temperatures, lifetimes, proper motions, spatial distribution and morphological evolution. Brightness of UDs  is confirmed to depend on the brightness of their surrounding  background. Several UDs show fission and fusion. Thanks to the stable condition of space observation, we could first follow the temporal behavior of these events. The derived properties of internal structure of the umbra are discussed in viewpoint of magnetoconvection in a strong magnetic field.
\end{abstract}

\section{Introduction}
Umbral fine structure in sunspots has been studied by many authors. Recent reviews were given in Thomas and Weiss (2004) and in references cited therein. The study of umbral fine features is very essential for our understanding of the magnetoconvection in a strong magnetic field atmosphere of celestial bodies. As the spatial size of umbral fine structure, such as umbral dots (UDs), was very fine, it was very hard to get the basic characteristics of them. Especially it was very difficult to follow the temporal evolution of the fine features from the ground-based telescopes, due to the influences of variable atmospheric seeing conditions.

Solar Optical Telescope (SOT) on board Hinode successfully launched on September 23, 2006 was designed to observe the solar fine structure with a 50 cm mirror from space (Kosugi et al. 2007; Tsuneta et al. 2007; Suematsu et al. 2007; Ichimoto et al. 2007; Shimizu et al. 2007). The resolving power in flight condition was confirmed to have nearly the theoretical one of 0.2 arcsec. With the Hinode/SOT, we observed the temporal evolution of umbral fine structures during the period of March 2-4, 2007. 
Spatial distribution of umbral structure, its temporal evolution, lifetimes, proper motions, temperatures were studied during the three days period. Besides the basic characteristics stated above, we could follow temporal evolution of fission and fusion events of umbral dots. In the following sections, we describe the details of observation and analysis procedures in Section 2, give our results in section 3, and finally discuss and summarize our results in section 4.

\section{Observation and Reduction}
\begin{figure}
  \begin{center}
    \FigureFile(60mm,60mm){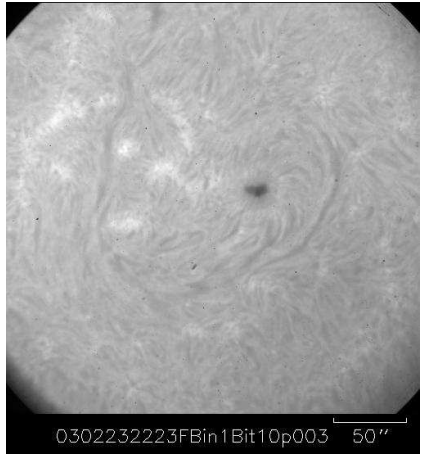}
    %%% \FigureFile(width,height){filename}
  \end{center}
  \caption{H$\alpha$ image of NOAA10944 on March 2, 2007 taken by DST at Hida Observatory.}\label{fig:kitai_fig1}
\end{figure}

\begin{figure}
  \begin{center}
    \FigureFile(80mm,240mm){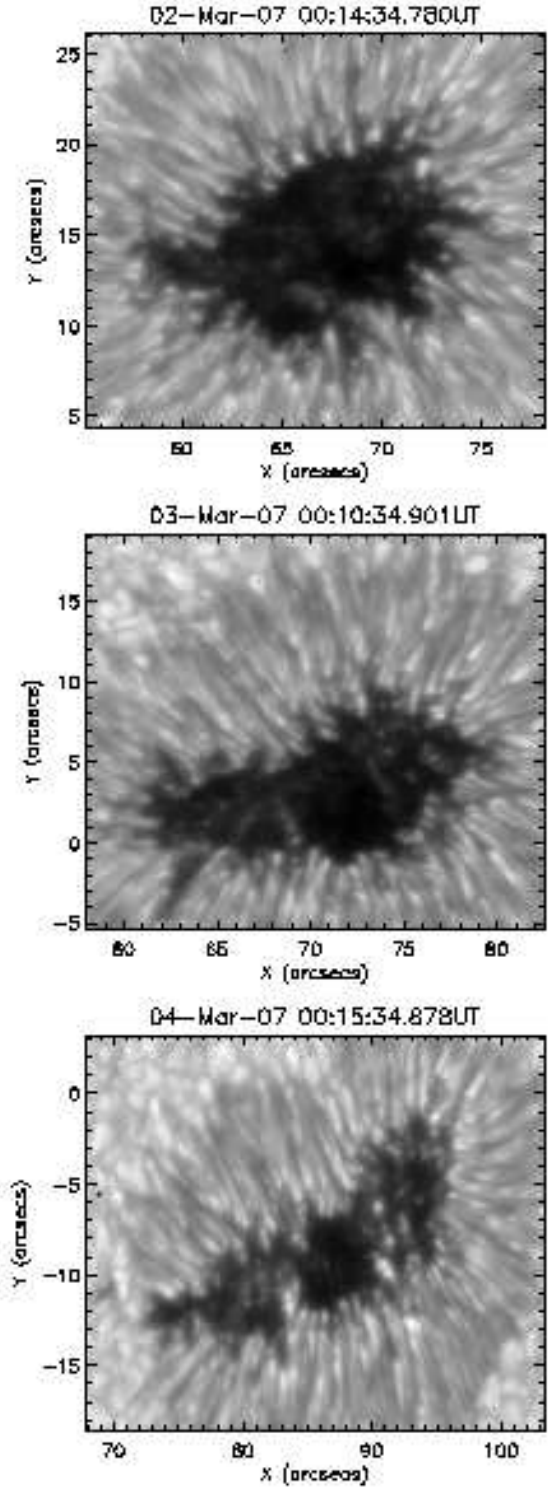}
    %%% \FigureFile(width,height){filename}
  \end{center}
  \caption{Daily evolution of the sunspot in green continuum.}\label{fig:kitai_fig2}
\end{figure}

We observed a roundish sunspot in an active region NOAA10944 from March 2 through March 4, 2007. The region was fairly inactive during the three days period and disintegrated on March 5. The region observed in H$\alpha$ with the Domeless Solar Telescope (DST) at Hida Observatory is shown in figure \ref{fig:kitai_fig1}. The main sunspot remained as $\alpha$ type during three days. Among the data taken with the Hinode/SOT, we will report the results obtained from time-series imaging observation by the Broadband Filter Imager (BFI), shown in table \ref{tab:kitai_tab1}. The green continuum images were taken through a filter ($\lambda=$ 5550\AA, $\Delta\lambda\simeq$ 5\AA), while the blue continuum images were through a filter ($\lambda=$ 4504\AA, $\Delta\lambda\simeq$ 5\AA). Both continuum images were taken in a cadence of 1frame/30sec. The pixel resolution of the images was \timeform{0''.054}. The field of view (FOV) of the continuum images was \timeform{55''.8}x \timeform{55''.8}. To follow the temporal evolution correctly without the projection effect, we transformed all the images as if they are seen from the top. Daily evolution of umbral region in green continuum is shown in figure \ref{fig:kitai_fig2}. 

\begin{figure}
  \begin{center}
    \FigureFile(60mm,200mm){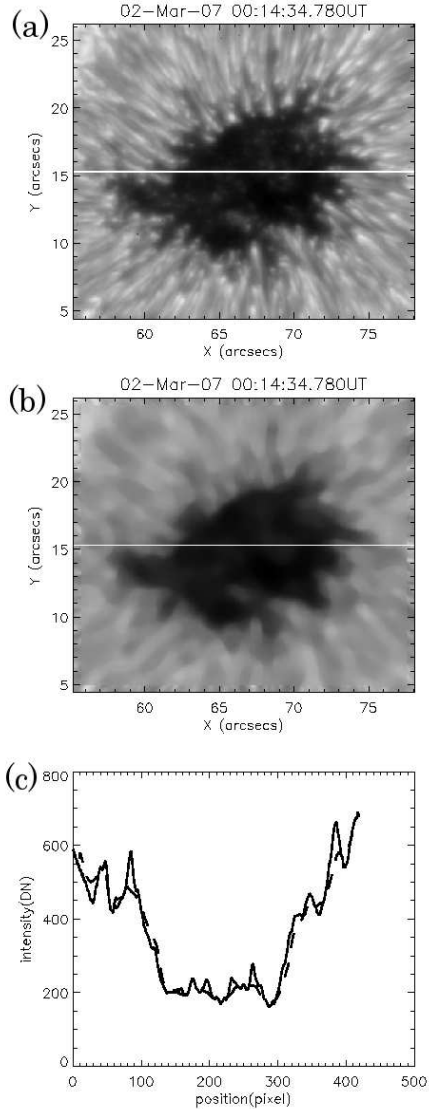}
    %%% \FigureFile(width,height){filename}
  \end{center}
  \caption{Effect of median filter. (a) Original image. (b) Filtered image. (c) Intensity plot along the white line indicated in (a) and (b). Both intensity profiles of original image (solid) and of median-filtered one (dashed) are shown. }\label{fig:kitai_fig3}
\end{figure}

We applied a median filter ( window : \timeform{1''} x \timeform{1''}  ) to all the images to identify slowly varying features, such as dark core area and diffuse components. The effect of median filter processing for structure identification is shown in figure \ref{fig:kitai_fig3}. All the images were co-aligned among them by finding image displacements which gives the maximum correlation between consecutive frames.

\begin{figure}
  \begin{center}
    \FigureFile(80mm,80mm){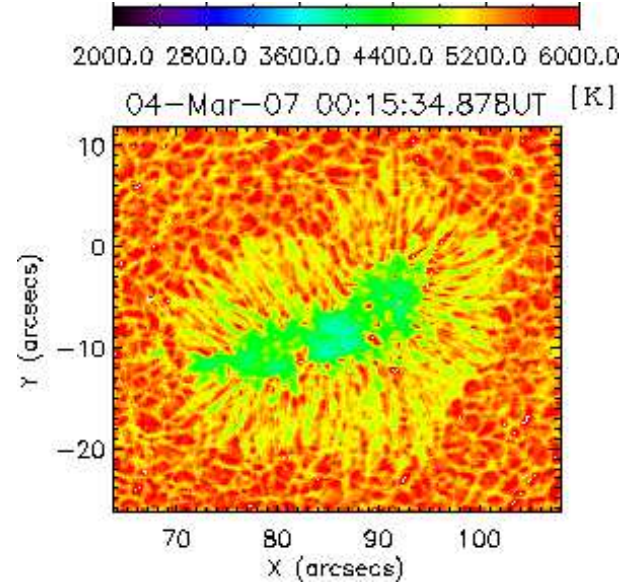}
    %%% \FigureFile(width,height){filename}
  \end{center}
  \caption{Temperature distribution on March 4, 2007.}\label{fig:kitai_fig4}
\end{figure}

Proper motions of UDs were derived by tracking the identified features along the time series of the images. Identification of features were done visually on PC screen. Lifetimes of UDs were determined by measuring the time spans, during which UDs showed 1.2 times larger brightness than their surrounding background. Temperatures of umbral features were estimated from color values , i.e., the intensity ratio I(blue)/I(green). The relation between the intensity ratio and temperature was calculated assuming the black body radiation. The temperature distribution over the region is shown in figure \ref{fig:kitai_fig4}. The temperatures of normal granules surrounding the spot are $\simeq$6000K, while those of intergranular lanes are $\simeq$5000K. These temperature values are consistent with those thus far known.

\begin{longtable}{lll}
  \caption{Observation}\label{tab:kitai_tab1}
  \hline 
  Date & Time & Filter \\ 
\endfirsthead
  \hline
\endhead
  \hline
%Date & Time & Filter \\ 
\endfoot
  \hline
\endlastfoot
  \hline
  2007 March 2 & 00:14-03:15 UT & green continuum \\
  2007 March 3 & 00:10-03:30 UT & green continuum \\
  2007 March 4 & 00:15-03:05 UT & green and blue continua \\
\end{longtable}

\section{Internal Structure of Umbral Region}
As is shown in figure \ref{fig:kitai_fig2}, brightness distribution of the umbral area is not uniform. The umbra observed by us consists of dark core region, diffuse components and bright umbral dots, as was observed in previous ground-based works under superb atmospheric conditions (Thomas and Weiss 2004). In our observation, the dark core almost kept its location and size, while the spot gradually evolved and deformed during three days period. In the dark core, UDs were very scarecely detected. Diffuse components were observed to stay at nearly the same location and develop into light bridges. UDs were numerously detected to appear and disappear except in the dark core region. Characteristics of individual components will be studied separately in the following subsections.
   
\subsection{Dark Core}
Daily evolution of umbral core is shown in figure \ref{fig:kitai_fig5}. Temperature of dark core is around 3850K on March 4. As the continuum brightness gradually increased, the core temperature is expected to have increased along the time. The tendency seems to be natural, as the spot was in decaying phase. We could identify virtually no UDs in the dark core. The absence or very low brightness of UDs in dark cores is confirmed in our observation from space, without the ambiguity of seeing conditions in ground-based observations. This is, probably, due to the positive correlation between UD's brightness and the background brightness, as will be stated in later section.

\subsection{Diffuse Component}
\begin{figure}
  \begin{center}
    \FigureFile(60mm,240mm){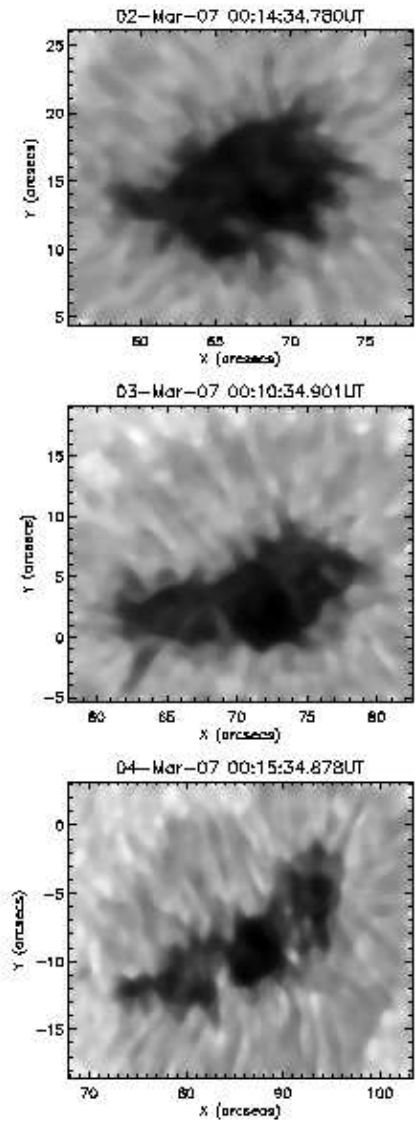}
    %%% \FigureFile(width,height){filename}
  \end{center}
  \caption{Daily evolution of dark core and diffuse components in green continuum. Median-filtered images are shown.}\label{fig:kitai_fig5}
\end{figure}

In figure \ref{fig:kitai_fig5}, we see daily evolution of diffuse components. Their locations were rather stable during the three days. They increased their brightness along time. Temperature ranged from 4250K to 4500K on March 4, which is 500K hotter than the dark core. They finally took the form of light bridges. In our case, the light bridges are of umbral type, neither of penumbral type nor of photospheric type, so the fine structures in the diffuse components have forms similar to UDs (Muller, 1979).

\subsection{Umbral Dots}
As were reported in many works (Kitai (1986), Sobotka et al.(1997a) and Thomas and Weiss(2004)), UDs are classified into two classes, i.e., (1)central or umbral origin, and (2) peripheral or penumbral origin. UDs of the former type appear and disappear in central parts of umbra. Their proper motions are known to be small. On the other hand, UDs of the latter type originate in the penumbral area. Tips of inner penumbral filaments start to be separated and move into umbral area with a larger velocity than the former type (Kitai, 1986). In our present study, we identified about 100 umbral dots during the three days period. 
We classified them, like previous works, into three classes according to their ways of birth, i.e., umbral origin (UUDs), penumbral origin (PUDs) and light-bridge origin (LUDs). 
\subsubsection{Size}
UDs generally have dot-like shapes. We measured linear sizes of them, and found that the sizes of majority of UDs are from \timeform{0''.32}(220km) to \timeform{0''.5}(350km). However, several percents of UDs have linear sizes of \timeform{0''.24}, i.e., the theoretical resolution limit of the telescope. We should have in mind that much smaller UDs can exist in umbrae. 
\subsubsection{Lifetime}
The lifetimes of UUDs range from 4 through 20 min and their average is 14.6 min. Those of PUDs range from 5 through 35 min and their average is 13.9 min, while LUDs ranges from 4 through 40 min, and their average is 16.0 min. So lifetimes of UDs do not depend on the types of UDs.
\subsubsection{Proper Motion}
The proper motions of both UUDs and LUDs showed similar behavior. Their speeds are virtually null, 0.5km s$^{-1}$ at maximum. Their directions of motions were random. On the other hand, PUDs showed higher speeds of about 0.9km s$^{-1}$ at their birth and gradually slowed down to 0.5km s$^{-1}$. 
\subsubsection{Temperature}
The temperatures of UUDs ranges from 4200K to 5500K, and their average is 4600K. Those of LUDs ranges from 4800K to 5600K, and their average is 5100K. PUDs are generally hotter at their birth and then become cooled down. Their temperatures range from 4700K to 5900K, and their average is 5460K. 
\subsubsection{Light Curve}
\begin{figure}
  \begin{center}
    \FigureFile(60mm,160mm){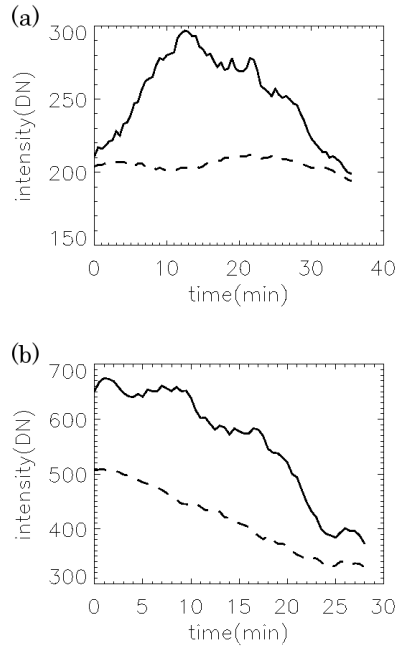}
    %%% \FigureFile(width,height){filename}
  \end{center}
  \caption{Light curve of a typical UD. (a) UUD/LUD. (b) PUD. Dashed line indicates the brightness just around the UD.}\label{fig:kitai_fig6}
\end{figure}

The temporal variation of brightness is found to depend on the type of UDs. UUDs and LUDs increase their brightness linearly, and then darken linearly along time. Light curve of a long-lived  UUD/LUD is shown in figure \ref{fig:kitai_fig6}(a). On the other hand, PUDs darken continuously  ( figure \ref{fig:kitai_fig6}(b) ). 

\subsection{Brightness/Temperature of UDs and surroundings}
\begin{figure}
  \begin{center}
    \FigureFile(80mm,80mm){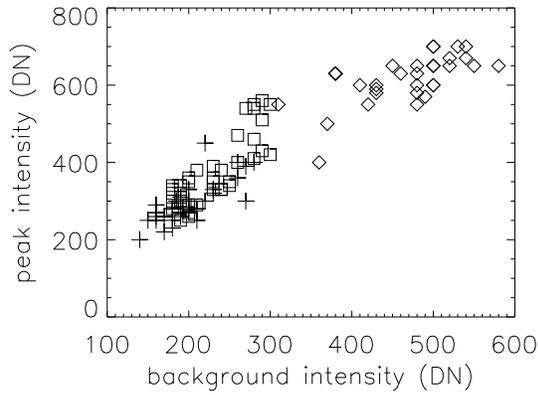}
    %%% \FigureFile(width,height){filename}
  \end{center}
  \caption{Scatter diagram of peak brightness of UDs against their background brightness. Plus-, square- and diamond-symbols are for UUD, LUD and PUD, respectively.}\label{fig:kitai_fig7}
\end{figure}

The brightness of UDs is found to depend on those of surrounding brightness. UDs seen in brighter background appear brighter/hotter than those in dark regions. Correlation between peak brightness of UDs and their background brightness is shown in figure \ref{fig:kitai_fig7}. From our temperature analysis, UDs are found to be around 300K hotter than their surroundings, irrespective of the type of UDs. The relation was first reported by Sobotka et al. (1992a), and have been studied by Sobotka et al. (1992b and 1993) and recently by Sobotka and Hanslmeier (2005). Our observation from space with \timeform{0''.24} resolution fully confirms their results, from the analysis of blue/green continuum brightness. 

\subsection{Fission and Fusion of Dots }

\begin{figure}
  \begin{center}
    \FigureFile(60mm,80mm){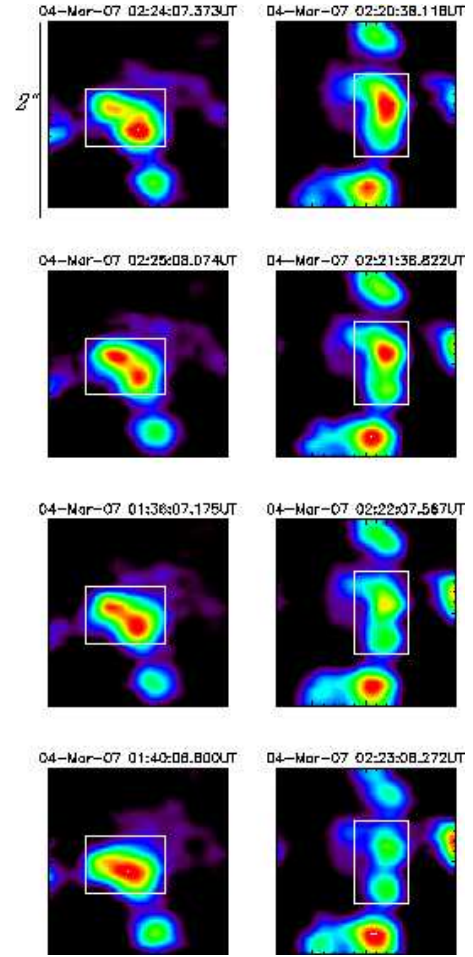}
    %%% \FigureFile(width,height){filename}
  \end{center}
  \caption{ Temporal evolution of fission and fusion of UDs shown in pseudo-color. Left column shows the fusion of two UDs, while right column shows the fission of another UD. From all the images, background intensities were subtracted to enhance the dot structures. }\label{fig:kitai_fig8}
\end{figure}

Some of UUDs and LUDs show fission and fusion, while majority of UDs keep their identities during their lives. One case of fission was observed in the sample of 30 UUDs, while one fission and two fusion events were detected in the sample of 31 LUDs. Sobotka et al. (1997a) noticed these events in tracking the evolution of UDs.  Temporal behaviors of fission and fusion are shown in figure \ref{fig:kitai_fig8}. Fissions occurred at the end of UDs's life.  They disintegrated into a smaller parts and faded away. On the other hand, two UDs merged into one and form a bright UD when fusion of UDs occurred. As for the PUDs, we have not detected these phenomena. The detailed evolution of the events was first observed in our work.
 
As UDs can be smaller than \timeform{0''.24} (Sobotka et al. 1997a; Sobotka and Hanslmeier 2005), fusion and fission events may be due to temporal brightness variation inside an unresolved cluster of  much smaller UDs. Fusion/fission may correspond to brightening/decaying phase of such a cluster of UDs.
     
\section{Summarizing Discussion}
Weiss et al. (2002), from their simulation of three-dimensional non-linear magnetoconvection in a strongly stratified compressible layer, gave two important suggestions on the internal structure of sunspots. First one was that UDs are manifestation of small scale convective cells, whose sizes are reduced strongly compared to normal granules by the strong magnetic field. Second one was that diffuse background components, which are larger in size than UDs, correspond to clumps of vigorously convecting plumes, from which magnetic flux is expelled. According to their suggestion, umbral area is separated into (a) regions of strong fields and small-scale convection and (b) regions of weak fields and large-scale vigorous convection.  Spectroscopic observations on the magnetohydrodynamic (MHD) nature of UDs have been done by several researchers. Wiehr and Degenhardt(1993) and  Socas-Navarro et al. (2004) got the results that the magnetic field strength is weaker and that small or virtually null upflows exist in UDs, while Lites et al (1991) detected no indication of reduced field strength in UDs. It seems that no comprehensive observational view of MHD behavior of UDs is obtained at present. Especially the differences of MHD behavior among the types of UUD/LUD/PUD are unknown.

 As was suggested in previous works and confirmed in our present observation, PUDs have different characteristics from UUDs/LUDs in view points of their birth places, and proper motions. As UUDs/LUDs are immobile and hotter than their surroundings, they probably correspond to small-scale magnetoconvection in a strong magnetic field, as was suggested by Weiss et al. (2002). On the other hand, PUDs show systematic proper motions from penumbrae to umbrae and seem to be natural extension of so called penumbral grains. We think that the interchange instability model at penumbrae by Schlichenmaier et al. (1998) well explains PUDs. However, one important point remains unexplained. UUDs/LUDs and PUDs have a common size of around 200-300km. Why do UUDs/LUDs of magnetic convective origin in umbrae and PUDs of penumbral origin have nearly the same geometrical sizes?  What are physical factors which determine the UD size of 200-300km? Are there a common mechanism which controls the sizes of fine structures both in umbrae and penumbrae?

Diffuse components are observed to be hotter than the dark core of umbrae. So they probably will correspond to plumes of vigorous convection with weak magnetic flux, as was suggested by Weiss et al. (2002). As LUDs observed in these diffuse components are observed to have much higher temperature than UUDs, we suspect that convection is stronger in these diffuse components.  However, according to Weiss et al. (2002), small-scale convective plumes, i.e. UDs, are not expected to appear in large-scale vigorous plumes of convection, i.e. diffuse components. This expectation is contrary to our observed result. In our three days observation, it was observed that diffuse components and the dark core kept their identities. Garcia de la Rosa (1987), from the analysis of temporal evolution of many sunspots, suggested that sunspots consist of a cluster of several large fragments and that these fragments keep life-long identities from the birth to the decay of sunspots. In the fragments, UUDs will be formed as was suggested by Weiss et al. (2002). At the interfaces of fragments, convection from deeper layers is expected to intrude the interfaces, resulting in rather bright diffuse components. 

In both the vigorous convective plumes model by Weiss et al.(2002) and the fragment model by Garcia de la Rosa (1987), the occurrence of LUDs in diffuse components remains unexplained. It may be related to the question why UDs have common geometrical sizes irrespective of their types. Similar mechanism such as the interchange instabilities proposed by Schlichenmaier et al. (1998) may work to form UDs in addition to magnetoconvetion.

Discussion and suggestions stated above, including the conjecture of the absence of UDs in the dark core (Lites et al., 1991), are to be studied by further observation and analysis. In our next paper, we will report our analysis of Spectropolarimeter data obtained on the same sunspot by Hinode/SOT. Temporal evolution of vector magnetic field and Doppler velocities in and around the UDs will give us more conclusive views of magnetoconvection in sunspots.

The authors acknowledge Drs. M. Sobotka and T. Yokoyama whose comments are valuable for the improvement of the paper. The authors are partially supported by a Grant-in-Aid for the 21st Century COE 'Center for Diversity and Universality in Physics' from the Ministry of Education, Culture, Sports, Science and Technology (MEXT) of Japan, and also partially supported by the Grant-in-Aid for 'Creative Scientific Research The Basic Study of Space Weather Prediction' (17GS0208, Head Investigator: K. Shibata) from the Ministry of Education, Science, Sports, Technology, and Culture of Japan. Hinode is a Japanese mission developed and launched by ISAS/JAXA, with NAOJ as domestic partner and NASA and STFC (UK) as international partners. It is operated by these agencies in co-operation with ESA and NSC (Norway). 

%%%%%%%%%%%%%%%%%%%%%%%%%%%%%%%%%%%%%%%

%%%
% See the manual for the detail.
%%%

\end{document}